\documentclass[aps,pra,twocolumn,floatfix,citeautoscript,nofootinbib,superscriptaddress]{revtex4-1}
\usepackage{amsbsy}
\usepackage{latexsym,epsfig,graphicx}
\usepackage{dcolumn}
\usepackage{graphicx}
\usepackage{subfigure}
\usepackage{comment}
\usepackage{color}
\usepackage{bm}
\usepackage{mathrsfs}
\usepackage{amsfonts}
\usepackage{amsmath}
\usepackage{color}
\usepackage{amssymb}
\usepackage{xspace}
\usepackage{epstopdf}
\usepackage{tabularx}
\usepackage{longtable}
\usepackage[colorlinks=true, letterpaper=true, pdfstartview=FitV, linkcolor=blue, citecolor=blue, urlcolor=blue]{hyperref}
\usepackage[normalem]{ulem}

\begin{document}

\title{Anti-helical edge magnons in patterned antiferromagnetic thin film}
\author{Yun-Mei Li}
\email{yunmeili@xmu.edu.cn}
\affiliation{Department of Physics, School of Physical Science and Technology, Xiamen University, Xiamen 361005, China}

\begin{abstract}
  Helical edge states in topological insulators give counterpropagating spin current on the two parallel edges.
  We here propose anti-helical edge states of magnons in patterned antiferromagnetic thin films,
  which host copropagating spin current on the two parallel edges, where the two magnon modes
  with opposite chirality act like the spin. The embedded heavy metal dot array
  in the thin film induces interfacial Dzyaloshinskii-Moriya interactions (iDMIs), drives
  the magnon bands into nontrivial topological phases, characterized by spin Chern number.
  The resulting helical edge modes lead to spin current with the direction dependent on the sign of iDMI parameter.
  In a strip geometry, we combine two subsystems with two embedded metal dot arrays, which give opposite
  iDMI parameters. Anti-helical edge states emerge, compensated by the counterpropagating bulk confined states.
  Helical and anti-helical edge states are verified by the micromagnetic simulations.
  Our work is quite helpful in the field of magnon spintronics based on antiferromagnets.
\end{abstract}

\maketitle

\section{Introduction}

Chiral or helical edge states are a hallmark feature of two-dimensional topological insulators~\cite{MZHasan,XLQi}.
Chiral edge states emerge in quantum Hall or Chern insulators with time-reversal symmetry breaking and propagate along the edges clockwise or counterclockwise.
As a consequence, a typical strip geometry supports opposite propagating edge states along the two parallel edges.
Helical edge states can be treated as two copies of chiral edge states,
protected by the time-reversal symmetry, holding great potential applications in spintronics.
The edge states for opposite spin hold opposite chirality, leading to a vanished charge current and a net spin current,
which flow in opposite direction at the two parallel edges in the strip geometry.

Recent theory predicted the so-called antichiral edge states~\cite{EColomes,DBhowmick,XCheng,YangY,PZhou,MMDenner,SMandal,JChen},
which copropagate in the same direction at opposite edges,
compensated by counterpropagating bulk modes, distinct from the chiral edge states.
The original work is based on the modified Haldane model in electronic systems~\cite{EColomes}, quite hard to realize experimentally.
Then a series of the theoretical even experimental works addressed the realization in many other systems, including
the electron system based on graphene~\cite{MMDenner,XCheng},
exciton polariton~\cite{SMandal}, magnon~\cite{DBhowmick}, photon~\cite{PZhou,JChen} and classical electric circuit~\cite{YangY} systems.
The experimental works in photonic~\cite{PZhou} and electric circuit~\cite{YangY} systems observed the
copropagating behavior of antichiral edge states based on the modified Haldane model.
All these works require the broken time-reversal symmetry.
A natural question arises. In the time-reversal symmetric systems, is it possible to realize ``anti-helical" edge states,
where the spin current on the parallel edges propagates in the same direction?
For now, quite few works made the explorations theoretically or experimentally.

In recent yeas, antiferromagnets have attracted significant attention in spintronics due to the ultrafast spin dynamics and lack of stray fields~\cite{TJungwirth,VBaltz}.
Importantly, antiferromagnets support both left-handed and right-handed polarized magnon modes, related by the pseudo-time reversal symmetry~\cite{YMLi1},
analogous to the electron spin. This coexistence of both polarizations give rise to many novel spin-related physical phenomena,
such as the magnon spin Nernst effect~\cite{RCheng,VAZyuzin}, Stern-Gerlach effect~\cite{ZWang} and Hanle effect~\cite{TWimmer}.
Manipulating the polarization may facilitate the chirality-based computing~\cite{CJia,MWDaniels} and logic devices~\cite{WYu1}.
Therefore, the exploration on topologically protected helical or even anti-helical edge state could be greatly helpful in the field of
magnon spintronics based on antiferromagnets~\cite{ABarman}.

In this paper, we theoretically propose the realization of anti-helical edge states of magnons
in patterned antiferromagnetic thin film, in which the spin current at parallel edges flow in the same direction,
compensated by the counterpropagating bulk confined spin currents, as illustrated in Fig.~\ref{fig1} (a).
Here the degenerate two magnon modes with opposite chirality is treated
as the spin degree of freedom, as widely discussed in previous works above.
Heavy metal dot array is embedded into the thin film to fold the free dispersion of magnons into bands.
Moreover, the inversion symmetry breaking at the interfaces between metal dots and the thin film will generate
interfacial Dzyaloshinskii-Moriya interactions (iDMIs),
which modify the exchange boundary condition and induce nontrivial topological phase for the magnon bands,
characterized by the spin Chern number. As a result, helical edge magnons arise in a finite width strip geometry.
The left-handed and right-handed edge magnons flow towards opposite
directions at the same edge, leading to a pure spin current, revealed by the micromagnetic simulations.
The direction of the spin current is determined by the sign of iDMI parameter $D$.
In a finite width ribbon geometry, we combine two subsystems embedded with
different metal dot arrays, which induce iDMIs with opposite sign of $D$, as shown in Fig.~\ref{fig1} (a).
Interesting, the spin currents along the two edges flow in the same direction, while the spin currents along the domain wall flow in opposite direction.
The emergence of the anti-helical edge states can also be verified by the micromagnetic simulations.

This paper is organized as follows. In Sec. II, we derive the eigenvalue problem for the magnons in the patterned antiferromagnetic thin film
in the presence of metal dot array and the iDMIs.
In Sec. III, we discuss the band topological properties due to the iDMIs, the topological phase transition and the construction of strip geometry for
anti-helical edge states. We also performed the micromagnetic simulations for verifications.
In Sec. IV, we summarize our results. In Sec. V, we present the methods and the parameters we adopted for the calculations.

\section{The patterned antiferromagnetic thin film}

\begin{figure}[t]
  \centering
  \includegraphics[width=0.45\textwidth]{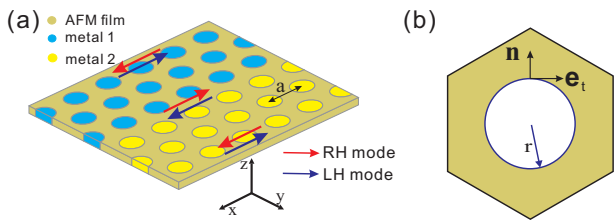}\\
  \caption{(a) The illustration of anti-helical edge spin waves in a patterned antiferromagnetic thin film.
  The heavy metal dot array is arranged in a triangular lattice configuration. We embedded two different metal dot array to construct the anti-helical
  edge states. The nearest distance between the dot center is $a$.
  (b) Top view of the unit cell for the antiferromagnetic thin film when only one type metal is embedded into the thin film.
  $\mathbf{n}$ and $\mathbf{e}_{t}$ are the normal and in-plane tangential vectors of the interfaces, respectively.
  The radius of the metal dot is denoted as $r$.}\label{fig1}
\end{figure}

We consider an antiferromagnetic thin film with periodically embedded metal dots.
The metal dots are arranged in a triangular lattice configuration.
We construct a domain along two sides with different embedded metal dot array
to get the anti-helical edge states in a strip geometry, as illustrated in Fig.~\ref{fig1} (a).
An infinite system with only one type embedded metal dot array is a two-dimensional artificial magnonic crystal.
The corresponding unit cell is shown in Fig.~\ref{fig1} (b).
The strip geometry with one or two embedded metal dot array is one-dimensional artificial magnonic crystal.
The magnetization dynamics in antiferromagnets can be described
by two coupled Landau-Lifshitz-Gilbert (LLG) equations for each sublattice,
\begin{equation}\label{eq1}
  \mathbf{\dot{m}}_{i}=-\gamma\mathbf{m}_{i}\times\mathbf{h}_{i}+\alpha\mathbf{m}_{i}\times\mathbf{\dot{m}}_{i},
\end{equation}
where $i=1,2$ denote the two sublattices. $\gamma$ is the gyromagnetic ratio, $\alpha$ is the Gilbert damping constant.
Here $\gamma\mathbf{h}_{i}=K_{z}m_{i}^{z}\mathbf{z}+A\nabla^{2}\mathbf{m}_{i}-J_{ex}\mathbf{m}_{\bar{i}}$
(with $\bar{1}=2$ and $\bar{2}=1$) is the effective magnetic field acting locally on sublattice $\mathbf{m}_{i}$,
where $K_{z}$ is the easy-axis anisotropy along $\mathbf{z}$ direction
$A$ and $J_{ex}$ characterize the Heisenberg exchange coupling constant of intra-sublattice and inter-sublattice, respectively.
The equilibrium magnetization is in collinear order for the two sublattices and parallel to the $z$-direction.
To get the magnon dispersions, we can divide the magnetization and effective magnetic fields on each sublattice
into the static and dynamical ones.
Let $\mathbf{m}_{i}=\mathbf{m}_{i}^{0}+\delta \mathbf{m}_{i}$ and $\mathbf{h}_{i}=\mathbf{h}_{i,0}+\delta\mathbf{h}_{i}$
with $\mathbf{m}_{i}^{0}$ the equilibrium magnetization and $|\delta \mathbf{m}_{i}|\ll \mathbf{m}_{i}^{0}$.
$\mathbf{h}_{i,0}$ is the $z$-component of the effective magnetic field.
By neglecting the damping first, we have
$\partial\delta\mathbf{m}_{i}/\partial t =-\gamma(\mathbf{m}_{i}^{0}\times\delta\mathbf{h}_{i}+\delta \mathbf{m}_{i}\times\mathbf{h}_{i,0})$.
To solve the dispersion in the artificial crystals,
we employ the ansatz $\delta\mathbf{m}_{i}=\delta\mathbf{m}_{i}^{\mathbf{k}}(\mathbf{r})e^{i(\mathbf{k}\cdot\mathbf{r}-\omega t)}$,
and after a transformation, we can get the eigenvalue equations, given by
\begin{equation}\label{eq2}
  \omega\psi_{\pm,\mathbf{k}} =\mp\sigma_{z}\hat{H}_{\mathbf{k}}\psi_{\pm,\mathbf{k}},
\end{equation}
where $\psi_{\pm,\mathbf{k}}=(\delta m_{1,\pm}^{\mathbf{k}},\delta m_{2,\pm}^{\mathbf{k}})^{T}$,
$\delta m_{i,\pm}^{\mathbf{k}}=\delta m_{i,x}^{\mathbf{k}}\pm i\delta m_{i,y}^{\mathbf{k}}$,
$\hat{H}_{\mathbf{k}}=[K_{z}+J_{ex}+(\nabla+i\mathbf{k})^{2}]\sigma_{0}+J_{ex}\sigma_{x}$.
$\sigma_{0}$ is $2\times2$ identity matrix, $\sigma_{x,y,z}$ are the Pauli matrices.
The solution for $\psi_{+,\mathbf{k}}$ denotes left-handed modes while  $\psi_{-,\mathbf{k}}$ is solution
for right-handed modes. As discussed in our recent work~\cite{YMLi1}, the two polarized modes are related by a pseudo-time reversal symmetry.
Note that the two polarized modes are totally decoupled.

We turn to the role of the embedded heavy metal dots.
The first effect is to fold the free dispersion into magnon bands due to the scattering by the periodic interfaces from the metals.
More importantly, as discussed in many theoretical~\cite{LUdvardi,JHMoon,MdRKAkanda,KYamamoto,MKostylev}
and verified in experimental works~\cite{JCho,NembachH,KZakeri,PFerriani,KDi},
the heavy metals (such as Pt, W, Au, Re, Ir, etc) with strong spin-orbit coupling can generate chiral interfacial Dzyaloshinskii-Moriya interactions (iDMI)
due to the inversion symmetry breaking at the magnets/heavy metal interfaces. In our system, the iDMI $\mathbf{h}_{\mathrm{DM}}$ is given by
$\gamma\mathbf{h}_{\mathrm{DM}}=D\mathbf{z}\times(\mathbf{e}_{t}\cdot\nabla)\mathbf{m}_{i}$
by comparing our bent surface to the previous plane surface~\cite{JHMoon,KDi}.
$\mathbf{e}_{t}=\mathbf{n}\times\mathbf{z}$ is the in-plane tangential vector of the interface and $\mathbf{n}$ the
normal vector of the interface, both depicted in Fig.~\ref{fig1} (b). $D$ may be either positive or negative,
depending on the metal materials~\cite{MdRKAkanda,KYamamoto}.
Without loss of generality, we assume $D$ is uniform at the bent surfaces.
The iDMI will give an interface torque~\cite{MKostylev}, change the exchange boundary condition into the formalism below,
\begin{equation}\label{eq3}
  \mathbf{n}\cdot\nabla\mathbf{m}_{i}+\frac{Dd}{A}\mathbf{z}\times(\mathbf{e}_{t}\cdot\nabla)\mathbf{m}_{i}=0,
\end{equation}
with $d$ the thickness of interface atomic layer and $D_{0}=Dd$ is a constant, only depending the materials. This constant $D_{0}$ value
accounts for the experimentally observed approximately inverse thickness dependence of spin wave frequency shift in ferromagnets/heavy metal heterostructures~\cite{JCho,NembachH}.

By solving the eigenvalue equation in Eq.~(\ref{eq2}) with the boundary condition in Eq.~(\ref{eq3}),
we can obtain the magnon dispersions in the $\mathbf{k}$-space.
The strip geometry illustrated in Fig.~\ref{fig1} (a) can also be calculated.
The results are shown in the subsequent section.

\begin{figure*}[t]
  \centering
  \includegraphics[width=0.95\textwidth]{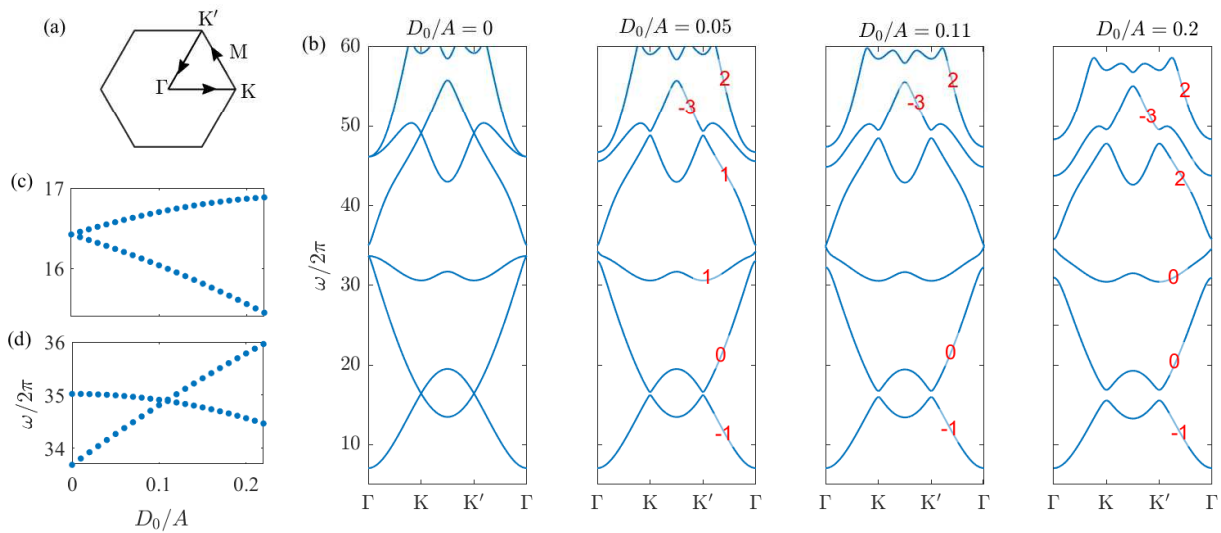}\\
  \caption{(a) The Brillouin zone and the path illustration for plotting the bands.
  (b). The band dispersion of right-handed magnon modes along high-symmetry directions illustrated
  in (a) at four different values of $D_{0}$: $D_{0}=0$, $D_{0}=0.05A$, $D_{0}=0.11A$ and $D_{0}/A=0.2A$, respectively.
  The left-handed mode are degenerate to the right-handed one.
  The corresponding Chern numbers of each band are indicated in red number. The Chern numbers for the left-handed modes are opposite.
  (c) The dependence of the eigenfrequency for the first and second bands at $\mathrm{K}$ ($\mathrm{K}^{\prime}$) point with respect to $D_{0}$.
  (d). The dependence of the eigenfrequency for the third and fourth bands at $\Gamma$ point with respect to $D_{0}$.}\label{fig2}
\end{figure*}

\section{Results and discussions}

\subsection{Band topological phase transitions}

Before presenting the construction of anti-helical edge states,
we first discuss the topological properties of magnons in AFM thin film embedded with only one type heavy metal dot array.
The corresponding Brillouin zone (BZ) is shown in Fig.~\ref{fig2} (a).
We present the band dispersions at four different $D_{0}$ values for right-handed modes in Fig.~\ref{fig2} (b).
The left-handed bands are degenerate to the right-handed ones.
When $D=0$, all the bands are topological trivial.
The first (fourth) and second (fifth) bands are degenerate at $\mathrm{K}$ and $\mathrm{K}^{\prime}$ points,
which gives multiple Dirac points in the spectrum.
The second (fifth) and third (sixth) bands intersect at $\Gamma$ point.

A finite $D$ lifts all the above the band degeneracies and open bandgap
at $\Gamma$, $\mathrm{K}$ and $\mathrm{K}^{\prime}$ points.
As the right-handed and left-handed modes are totally decoupled,
we can use the Chern number to characterize the topological properties for the bands of each polarized mode.
The berry connection of $n$-th band is given by
 $\mathbf{A}_{n,\pm}^{\mathbf{k}}=\mp i\int_{uc}\psi_{\pm,\mathbf{k}}^{n,\dagger}\sigma_{z} \nabla_{\mathbf{k}}\psi_{\pm,\mathbf{k}}^{n}d\mathbf{r}$
and Berry curvature $\mathbf{B}_{n,\pm}^{\mathbf{k}}=\nabla_{\mathbf{k}}\times \mathbf{A}_{n,\pm}^{\mathbf{k}}$.
The Chern number $C_{n,\pm}=\frac{1}{2\pi}\int_{BZ}B_{n,z,\pm}^{\mathbf{k}}d^{2}\mathbf{k}$.
The integral for the berry connection is done over the unit cell.
The Chern number for the corresponding right-handed magnon bands are shown in Fig.~\ref{fig2} (b) by the number on the bands.
We find the Chern numbers for the left-handed magnon bands are opposite to the right-handed ones for each degenerate band.
Therefore, we can use the spin Chern number to characterize the topological properties of the total system,
$C_{n,s}=(C_{n,-}-C_{n,+})/2$.
We can also define the $n$-th gap spin Chern number $\bar{C}_{n,s}$ associated with the gap
above the $n$-th band as
\begin{equation}\label{eq4}
  \bar{C}_{n,s}=\sum_{n^{\prime}=1}^{n}C_{n^{\prime},s}.
\end{equation}
The number $\bar{C}_{n,s}$ determined the number of helical edge states in the $n$-th gap according
to the bulk-boundary correspondence.
When $D$ reverse sign, the Chern number for each mode will reverse the sign accordingly.
The spin Chern number will also reverse sign. Therefore, the spin current due to the helical edge states are expected
to change the flow direction. As the sign of $D$ is determined by the heavy metal materials, we can design our system
by choosing the heavy metal as desired. Based on this feature, we can
construct the anti-helical edge magnons by structural design.

For the first gap, a larger $D$ will give a larger bandgap at $\mathrm{K}$ and $\mathrm{K}^{\prime}$ points,
as shown in Fig.~\ref{fig2} (c). At any value of $D$, the first gap spin Chern number is always $-1$.
Interestingly, increasing $D$ will drive a topological phase transition for the third and fourth bands.
We find when $D_{0}=D_{0}^{c}\simeq 0.11A$, the bandgap between the third and fourth bands at $\Gamma$ point will be closed.
A larger $D_{0}$ will reopen the gap, as shown in Fig.~\ref{fig2} (b) and (d). The spin Chern number
for the third band $C_{3,s}$ will change from $1$ to $0$, while for the fourth band, $C_{4,s}$ changes
from $1$ to $2$. The total (spin) Chern number of the two bands remain unchanged.
But this topological phase transition will change the third gap spin Chern number $\bar{C}_{3,s}$ from $0$ to $-1$.
This indicates that the third gap will experience a topologically trivial to nontrivial phase transition.
From Fig.~\ref{fig2} (b), we can see only the first and third bandgap are full gap.

\subsection{Helical edge magnons}

\begin{figure}[t]
  \centering
  \includegraphics[width=0.5\textwidth]{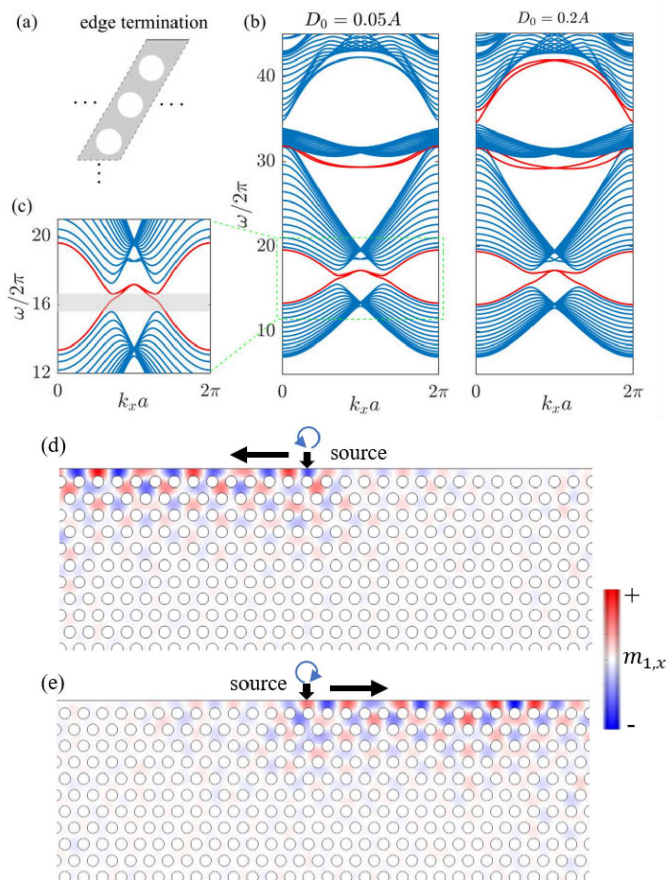}\\
  \caption{(a) The illustration of the edge termination on the upper edge for the finite width ribbon when only one heavy metal are embedded into the thin film.
  (b) The right-handed magnon band dispersions for the ribbon at $D_{0}=0.05A$ and $D_{0}=0.2A$, respectively. The left-handed magnon bands are degenerate
  to the right-handed ones. The edge states are displayed in red lines. The width of the ribbon is 716 nm.
  (c) The zoom-in bands near the first bandgap for $D_{0}=0.05A$ from the green dashed box shown in (b).
  (d) Snapshot of the propagation of the excited right-handed spin waves at frequency $\omega/2\pi=16.2$ GHz. The antenna position is show by the arrow.
  (e) Snapshot of the propagation of the excited left-handed spin waves at frequency $\omega/2\pi=16.2$ GHz. In (c) and (d), $D_{0}=0.2A$.
  The black arrow show the position of microwave antenna.}\label{fig3}
\end{figure}

The nonvanishing $n$-th gap spin Chern number is expected to give rise to helical edge states covering the bandgap.
To verify this prediction, we calculate the band dispersion in a strip geometry with a finite width along $y$ direction and infinite along $x$ direction.
The strip geometry with only the upper edge termination are shown in Fig.,~\ref{fig3} (a).
The ribbon band dispersions for the right-handed modes are shown in Fig.~\ref{fig3} (b) with $D_{0}=0.05A$ (left) and $D_{0}=0.2A$ (right).
The bands for left-handed modes are degenerate to the right-handed ones.
For $D_{0}=0.05A$, We can see the edge states cover the first and second bandgap.
The spin Chern number of the third gap is $0$, indicating a trivial gap without supporting edge states.
Edge states are also expected to exist in higher bands.
But due to the absence of full gaps, they are not distinguishable from the bulk states, thus  not shown.
For $D_{0}=0.2A$, the edge states in the first and second gaps preserve. Additional edge states emerge
in the third bandgap due to the topological phase transition
for $D_{0}>D_{0}^{c}$ compared to $D_{0}<D_{0}^{c}$.

From the zoom in bands for $D_{0}=0.05A$ near the first bandgap in Fig.~\ref{fig3} (c),
we can see that the edge states with the frequency located in the shaded region show
a totally chiral feature. These edge states thus propagate unidirectionally.
We can also see this feature for the states in the first and third gap for $D_{0}=0.2A$.
The excitation of the states with other frequencies in the first gap and the states in the second gap will do not show
unidirectional flow. In the subsequent simulations, we mainly focus on the states in the first bandgap.
Due to the helical feature of the edge states, pure spin current locate at the edges.
To verify the helical feature of topologically protected edge magnons,
we perform micromagnetic simulations by selectively exciting spin waves with either one polarization.
The method for the micromagnetic simulations is present in Sec V.
The results are shown in Fig.~\ref{fig3} (d) and (e).
We apply a right-handed microwave source with the frequency located in the first bandgap.
The excited right-handed spin waves flow unidirectionally towards the left [Fig.~\ref{fig3} (d)], consistent with the in-pane magnetization
distributions obtained from band calculation.
When we changed the microwave source to the left-handed one, the excited spin waves move towards right [Fig.~\ref{fig3} (e)].
When the excitation microwave source is linearly polarized, both magnon modes will be excited.
A higher frequency located in the third bandgap will also give the same behavior.
When the excitation source is placed in the opposite parallel edge, the moving direction of each mode will be opposite.
These behaviors of excited spin waves clearly demonstrated the helical feature of the edge states.
Same to the electronic systems, it will lead to a pure magnon spin current on each edge, counterpropagating along the two parallel edges.
These states are topologically protected and are thus robust against the disorders from the nanofabrication process,
same to our previous work~\cite{YMLi}.

As the sign of spin Chern number is locked to the sign of iDMI parameter $D$, which is dependent on the heavy metal materials.
When we embed different heavy metal dot array to induce an opposite $D$, both of the polarized edge states will reverse the propagation direction,
so as to the spin current. The feature in our system is the key point for the construction of anti-helical edge states.

\subsection{Anti-helical edge magnons}

\begin{figure}[t]
  \centering
  \includegraphics[width=0.45\textwidth]{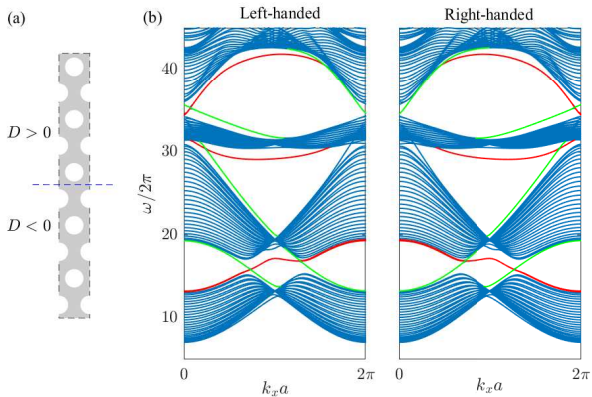}\\
  \caption{(a) illustration of  a ribbon geometry to support the anti-helical edge states. Above and below the blue dashed line, we adopt two different embedded heavy metal dots to
  induce opposite $D$ for the iDMI. (b) The band dispersion for the left-handed and right-handed modes for the configuration shown in (a).
  The width of the of the ribbon is 1236 nm. The edge states are displayed in red lines
  while the states confined on the interface formed by two metal dot array are displayed in green lines.
  The edge states are doubly degenerate localized on the upper and lower edges, respectively.
  The absolute value of $D_{0}$ for the calculation is both $0.2A$.
  }\label{fig4}
\end{figure}

Based on the locking behavior between spin Chern number and $D$, we now construct the anti-helical edge magnons in our system
to make the edge spin currents on the two parallel edges flow towards the same direction.
As the the sign of $D$ is determined by the heavy metal materials, we design an interface in the middle of the finite width strip, illustrated in Fig.~\ref{fig1} (a).
Two different heavy metal materials are embedded along the two sides of the interface.
The two metal materials are assumed to induce iDMI with opposite $D$, as illustrated
in Fig.~\ref{fig4} (a). For simplicity, we assume the absolute value of $D$ is the same as different values give the same physics.

We plotted the bands for both polarized modes in Fig.~\ref{fig4} (b) at $|D_{0}|=0.2A$.
The edge states are marked by red lines while the green lines denote the modes localized
along the interface between the two domains with different embedded heavy metal dot array.
These edge and interfacial states are still topologically protected.
The edge modes are doubly degenerate for both spin, localized on the upper and lower edges, respectively.
For each polarized edge states covering the first and third gaps, the two degenerate states propagate along the same direction,
while the the interfacial states flow towards the opposite direction.
The left-handed polarized bands can be mirrored to the right-handed ones respect to $k_{x}a=\pi$ (time-reversal invariant point) and vice versa.
On each edge and on the interface, the two polarized modes still counterflow to each other.
As a result, the spin current on the two edges flow towards the same direction,
compensated by the counterpropagating interfacial spin current, manifestation of the anti-helical feature for magnons
in this designed structure.

\begin{figure*}[ht]
  \centering
  \includegraphics[width=0.9\textwidth]{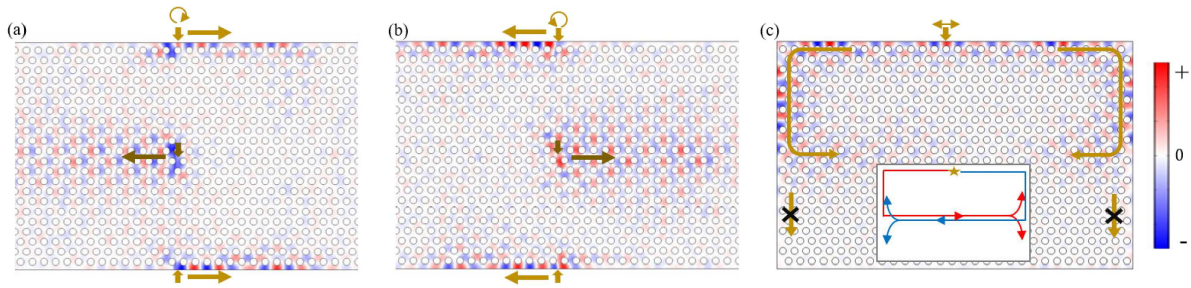}\\
  \caption{Snapshots of the spin wave propagations when microwave antennas is applied. The excitation frequency is $\omega/2\pi=16.2$ GHz.
   (a) Spin wave propagation of excited left-handed modes near the places marked by the arrows.
   The left-handed edge modes move towards to the right while the confined modes to the left.
   (b) Sin wave propagation of excited right-handed modes near the places marked by the arrows.
   The right-handed edge modes move towards to the left while the confined modes to the right, opposite
   to the left-handed modes. The colorbar for (a) and (b) is shown for the $m_{1,x}$.
   (c) Spin wave propagation when an linearly polarized microwave source is applied. Both left- and right-handed modes are excited.
   We marked the propagation path. In the inset, we give the evolution path of the both modes after a longer time with red (blue) line denoting the right-handed (left-handed) modes.
   The colorbar for (c) is for $n_{y}=m_{1,y}-m_{2,y}$. The cell size for the three figures is 1500 nm$\times$ 975 nm.}\label{fig5}
\end{figure*}

The anti-helical feature of the edge magnons can also be revealed in the micromagnetic simulations.
The results are shown in Fig.~\ref{fig5} (a) and (b).
The excited left-handed spin waves on the parallel edges flow towards right while the interfacial ones flow to the left.
On the contrary, the excited right-handed edge spin waves propagate towards left and the interfacial ones propagate to the right.
These behaviors demonstrate the successful construction of anti-helical edge magnons.
In a cell with a finite width along the $x$ direction, a linearly polarized microwave source will excite
both polarized spin waves, as shown in Fig.~\ref{fig5} (c). When the spin waves meet the left and right vertical edges,
they will be totally transmitted without backscattering due to the topological protection.
But they can not penetrate into the vertical edges lower than the interface.
This is because of the anti-helical feature, either polarized vertical edge magnons above and below the interface move in the opposite directions,
different from the chiral and helical edge states in topological insulators.
Instead, they will be transmitted into the interface, as shown in Fig.~\ref{fig5} (c).
However, after a longer time when transport to the opposite vertical edges, they will split into two beams, as illustrated
in the inset of Fig.~\ref{fig5} (c). One of the two will be transmitted to the lower horizontal edge.
All the edge magnons on each horizontal edge can be transmitted to the other only through the interface.

\subsection{Discussions}

Above all, we have discussed the band topology due to the iDMIs,
the helical also the anti-helical edge magnons in the patterned antiferromagnetic thin film. In our calculations, the thickness
of the film is 4 nm. The other thickness will also give the same band dispersions when only considering the $k_{z}=0$ bulk modes.
The modes with higher frequency at nonvanishing $k_{z}$ for any film thickness
are also expected to give the same physics as any $k_{z}$ mode share the same symmetry in the $xy$ plane.
We adopt the triangular configuration for the embedded heavy metal array.
The other configurations like the square lattice are also expected to give similar results.

We here adopted a medium lattice constant ($a=50$ nm) for the calculations.
A larger lattice constant will give a smaller bandgap,
but does not alter the topological protection of the edge and interfacial states.
Our model is experimentally feasible to construct.
The helical, anti-helical and interfacial states can be probed in the spin pumping process
as opposite magnon chirality could generate opposite spin polarization, discussed in recent experimental work~\cite{YLiu}.
They can also be detected by the polarization selective spectroscopy~\cite{YShiota}.
In addition, our configuration in a ferromagnetic thin film will lead to the antichiral edge states as there is only one magnon polarization in ferromagnets.

\section{Summary}

In this work, we proposed a theoretical scheme to realize the anti-helical edge magnons in antiferromagnetic thin film with embedded
heavy metal dot array. The iDMIs induce helical edge states with the direction of spin current dependent on the sign of iDMI parameter.
By constructing an interface with embedded different metal dots along the two sides in a finite width ribbon, edge spin current
on the parallel two edges propagate in the same direction, compensated by conterpropagating spin current along the interface.
Our proposal is experimentally feasible and the anti-helical edge states can be probed by the state-of-the-art techniques.
Our work provides a new method controlling the magnon spin in antiferromagnets and should be quite helpful in the field of magnon spintronics based on antiferromagnets.

\section{Methods}

The nearest distance between the metal dot center (lattice constant) is $a=50$ nm. The radius of the metal dot is $r=15$ nm.
The parameters in the coupled LLG equation are as follows:~\cite{WYu2}
$K_{z}=8.55$ GHz, $A=7.25\times10^{-6}$ Hz$\cdot$m$^{2}$, $J_{ex}=1.105\times10^{11}$ Hz.
The saturation magnetization $M_{s}=1.94\times10^{5}$ A/m,
the gyromagnetic ratio $\gamma=2.21\times10^{5}$ Hz$\cdot$m/A.
The values of $D_{0}$ in the unit of $A$ are given in the main text according to our need for the discussions.
The micromagnetic simulation is based on the Eq.~(\ref{eq1}). The damping constant is adopted as $\alpha=2\times10^{-4}$
to get a long decay length. We did not take the dipolar fields into calculation due to the antiferromagnet environment.

We can selectively excite either one or both of the polarized modes simultaneously by setting the microwave fields.
The effective field of the radio wave is given by $\mathbf{h}_{rf}(t)=h_{0}[\cos(\omega t)\mathbf{e}_{x}\pm\sin(\omega t)\mathbf{e}_{y}]$, where
$+$ denotes the right-handed source and $-$ denotes the left-handed source.
Only one component above along the $x$ or $y$ direction give a linearly polarized source.
In our micromagnetic simulation, $h_{0}=2\times10^{9}$ Hz, $\omega=2\pi\times16.2$ GHz.

\section{Acknowledgement}

This work is supported by the startup funding from Xiamen University.

\end{document}